\documentclass[twoside]{amsart}

\begin{document}

\title{Reformulating the Kriging Algorithm to the Practicing Miner}

\author{T. Sus{\l}o}
\email{tomasz.suslo@gmail.com}

\begin{abstract}
The main aim of applied geostatistics is to derive 
`mean' and `variance' of ore to the practicing miner. This paper suggests 
brand-new approach to the problem.
\end{abstract}

\maketitle

\thispagestyle{empty}

\section{Theory of weighted average}

At the base of the theory of weighted average we got the arithmetic mean
$$
\hat{m}=\omega'{\bf v}=\frac{1}{n}\sum^{n}_{i=1}v_i
$$
the simplest case of the ordinary least-squares estimator of mean value
$$
\hat{m}=\omega'{\bf v}=\frac{F'X^{-1}{\bf v}}{F'X^{-1}F} 
$$
a special case of the generalized least-squares estimator of mean value
$$
\hat{m}=\omega'{\bf v}=\frac{F'\Lambda^{-1}{\bf v}}{F'\Lambda^{-1}F}
$$
an asymptotic disjunction of the numerical generalized
least-squares estimator of mean value
$$
\hat{m}=\omega' {\bf v}
$$
if from the so-called \emph{kriging variance} holds
\begin{equation}
\omega' r + \mu = 0 \ ,
\label{ce}
\end{equation}
where (kriging algorithm)
\begin{equation}
\begin{array}{cccccl}
{\underbrace{
\left[
\begin{array}{cccc}
\Lambda &  F \\
F' & 0 \\
\end{array}
\right]}_{(n+1)\times (n+1)}}
& 
\cdot 
&
\underbrace{
\left[
\begin{array}{c}
\omega \\
\mu \\
\end{array}
\right]
}_{n+1\times 1}
&
=
&
\underbrace{
\left[
\begin{array}{c}
r \\
1 \\
\end{array} 
\right] 
}_{n+1 \times 1}   \ ,
\end{array}
\label{ke}
\end{equation}
$\Lambda$ is an auto-correlation matrix and $r$ is a correlation vector between known values at theirs co-ordinates and unknown value at desired co-ordinate.

\vspace*{12pt}
\noindent
{\bf Proof.} 
The ordinary least-squares estimator 
$$
\omega'{\bf v}=\frac{F'X^{-1}{\bf v}}{F'X^{-1}F} \ , 
$$
if $F'=[1,\ldots,1]$, where $X$ is an indentity matrix,
simplifies to the arithmetic mean
$$
\omega'{\bf v}=\frac{F' {\bf v}}{n}
$$
the generalized least-squares estimator
$$
\omega'{\bf v}=\frac{F'\Lambda^{-1}{\bf v}}{F'\Lambda^{-1}F}
$$
was generalized from the ordinary least-squares estimator
$$
\omega'{\bf v}=\frac{F'X^{-1}{\bf v}}{F'X^{-1}F} 
$$
since at long distance holds
$$
r=\xi F
$$
then from~(\ref{ce})
$$
\omega' r + \mu = \xi \omega' F + \mu = \xi + \mu = 0
$$
we get
$$
\mu = -\xi
$$
and~(\ref{ke}) 
$$
\begin{array}{cccccl}
{\underbrace{
\left[
\begin{array}{cccc}
\Lambda &  F \\
F' & 0 \\
\end{array}
\right]}_{(n+1)\times (n+1)}}
& 
\cdot 
&
\underbrace{
\left[
\begin{array}{c}
\omega \\
-\xi \\
\end{array}
\right]
}_{(n+1)\times 1}
&
=
&
\underbrace{
\left[
\begin{array}{c}
\xi F \\
1 \\
\end{array}
\right] 
}_{(n+1) \times 1} 
\end{array}
$$
with the least-squares solution 
$$
\omega = \frac{\Lambda^{-1} F}{F' \Lambda^{-1} F}
$$
for the estimation problem of mean value
$$
\omega'{\bf v}=\frac{ F' \Lambda^{-1} {\bf v}}{F' \Lambda^{-1} F} \ .
$$

\section{The best linear unbiased generalized statistics
of an unknown mean and variance value}
\noindent

The numerical generalized least-squares estimator
of mean value is
$$
\hat{m}= \omega' {\bf v}
$$
the numerical generalized least-squares estimator
of variance value is
$$
\hat{\sigma}^2=\omega' {\bf v}^2 - \hat{m}^2
$$
with the kriging weights given by the numerical approximation to the root of the equation
$$
\omega' r + \mu =0 \ ,
$$
where
$$
\begin{array}{cccccl}
{\underbrace{
\left[
\begin{array}{cccc}
\Lambda &  F \\
F' & 0 \\
\end{array}
\right]}_{(n+1)\times (n+1)}}
& 
\cdot 
&
\underbrace{
\left[
\begin{array}{c}
\omega \\
\mu \\
\end{array}
\right]
}_{(n+1)\times 1}
&
=
&
\underbrace{
\left[
\begin{array}{c}
r \\
1 \\
\end{array}
\right] 
}_{(n+1) \times 1}  
\end{array}
$$
with a mean squared error of mean estimation
$$
MSE(\hat{m})=\hat{\sigma}^2(\omega' r - \mu) \ .  
$$

\vspace*{12pt}
\noindent
{\bf Example.} Let us consider a stationary random field based on
coal seam thickness measurements taken over an approximately square area (Tab.\ref{Tab}) 
\begin{table}[!ht]
\small
\centerline{
\begin{tabular}{|c c c | c c c | c c c| }
\hline 
       0.7 & 59.6 & 34.1 & 2.1 & 82.7 & 42.2 & 4.7 & 75.1 & 39.5 \\ 
       4.8 & 52.8 & 34.3 &  5.9 & 67.1 & 37.0 &  6.0 & 35.7 & 35.9 \\
       6.4 & 33.7 & 36.4 &  7.0 & 46.7 & 34.6 &  8.2 & 40.1 & 35.4 \\  
      13.3 &  0.6 & 44.7 & 13.3 & 68.2 & 37.8 & 13.4 & 31.3 & 37.8 \\
      17.8 &  6.9 & 43.9 & 20.1 & 66.3 & 37.7 & 22.7 & 87.6 & 42.8 \\
      23.0 & 93.9 & 43.6 & 24.3 & 73.0 & 39.3 & 24.8 & 15.1 & 42.3 \\
      24.8 & 26.3 & 39.7 & 26.4 & 58.0 & 36.9 & 26.9 & 65.0 & 37.8 \\
      27.7 & 83.3 & 41.8 & 27.9 & 90.8 & 43.3 & 29.1 & 47.9 & 36.7 \\
      29.5 & 89.4 & 43.0 & 30.1 &  6.1 & 43.6 & 30.8 & 12.1 & 42.8 \\
      32.7 & 40.2 & 37.5 & 34.8 &  8.1 & 43.3 & 35.3 & 32.0 & 38.8 \\
      37.0 & 70.3 & 39.2 & 38.2 & 77.9 & 40.7 & 38.9 & 23.3 & 40.5 \\
      39.4 & 82.5 & 41.4 & 43.0 &  4.7 & 43.3 & 43.7 &  7.6 & 43.1 \\
      46.4 & 84.1 & 41.5 & 46.7 & 10.6 & 42.6 & 49.9 & 22.1 & 40.7 \\
      51.0 & 88.8 & 42.0 & 52.8 & 68.9 & 39.3 & 52.9 & 32.7 & 39.2 \\
      55.5 & 92.9 & 42.2 & 56.0 &  1.6 & 42.7 & 60.6 & 75.2 & 40.1 \\
      62.1 & 26.6 & 40.1 & 63.0 & 12.7 & 41.8 & 69.0 & 75.6 & 40.1 \\
      70.5 & 83.7 & 40.9 & 70.9 & 11.0 & 41.7 & 71.5 & 29.5 & 39.8 \\
      78.1 & 45.5 & 38.7 & 78.2 &  9.1 & 41.7 & 78.4 & 20.0 & 40.8 \\
      80.5 & 55.9 & 38.7 & 81.1 & 51.0 & 38.6 & 83.8 &  7.9 & 41.6 \\
      84.5 & 11.0 & 41.5 & 85.2 & 67.3 & 39.4 & 85.5 & 73.0 & 39.8 \\
      86.7 & 70.4 & 39.6 & 87.2 & 55.7 & 38.8 & 88.1 &  0.0 & 41.6 \\
      88.4 & 12.1 & 41.3 & 88.4 & 99.6 & 41.2 & 88.8 & 82.9 & 40.5 \\
      88.9 &  6.2 & 41.5 & 90.6 &  7.0 & 41.5 & 90.7 & 49.6 & 38.9 \\
      91.5 & 55.4 & 39.0 & 92.9 & 46.8 & 39.1 & 93.4 & 70.9 & 39.7 \\
      94.8 & 71.5 & 39.7 & 96.2 & 84.3 & 40.3 & 98.2 & 58.2 & 39.5 \\
\hline
\end{tabular}}
\caption{\label{Tab} East, north, thick.}
\end{table}     
with the theoretical correlogram given by 
$$
\rho(|h|)=\left\{
        \begin{array}{ll}
        +1 \cdot {\exp}^{-3(|h|\slash 30)^2},& 
        \qquad \mbox{for}~~|h|>0,\\
        +1, & \qquad \mbox{for}~~|h|= 0,\\
        \end{array}
        \right. 
$$
solving the kriging algorithm in the terms of correlation function for a grid: easting $-50$ to $+50$ and norting $-50$ to $+50$ 
with the step $0.1$ the minimum of the term
$$
|\omega' r + \mu|
$$
is reached at the node $-21.8$ east and  $42.6$ north
then the kriging weights give
$$
\hat{m}=\omega' {\bf v}=38.9 
$$
and
$$
\hat{\sigma}^2=\omega' {\bf v^2} - \hat{m}^2=16.1
$$
with
$$
MSE(\hat{m})=1.8
$$
whilist
$$
\hat{m}=\omega' {\bf v}=36.6
$$
and
$$
\hat{\sigma}^2=\omega' {\bf v^2} - \hat{m}^2 < 0
$$
if (the least-squares weights)
$$
\omega' = \frac{F' \Lambda^{-1}}{F' \Lambda^{-1} F} \ .
$$

\end{document}